\documentclass[11pt]{article}

\usepackage[margin=1in]{geometry}
\usepackage{amsmath,amssymb,amsfonts}
\usepackage{array}
\usepackage{booktabs}
\usepackage{hyperref}
\usepackage{authblk}

\title{Complex Variables for Fractional-Order Systems}
\author{Emmanuel A.\ Gonzalez}
\affil{E-mail: \texttt{emmgon@gmail.com}}
\date{August 30, 2018}

\begin{document}

\maketitle

\begin{abstract}
This paper discusses and summarizes some results on complex variables that are
very useful in fractional-order systems analysis and design, specifically when the system
is analyzed in the frequency domain. The author hopes that this document will serve
as a handy reference when performing computations with complex variables, especially
when working within the Laplace or Fourier domains. The reader can refer to Table~\ref{tab:summary}
for the summary of these formulas.
\end{abstract}

%--------------------------------------------------------------------------
\section{Basics}
%--------------------------------------------------------------------------

A complex variable
\[
  s = \sigma + j\omega
\]
has two parts: the real part represented by the variable~$\sigma$ and an imaginary part
represented by the variable~$\omega$. The real part of~$s$ is denoted by $\operatorname{Re}(s)$
such that $\operatorname{Re}(s)=\sigma$, while the imaginary part of~$s$ is denoted by
$\operatorname{Im}(s)$ such that $\operatorname{Im}(s)=\omega$.

Complex variables can also be added and multiplied. Consider two complex variables
$s_1=\sigma_1+j\omega_1$ and $s_2=\sigma_2+j\omega_2$. The following complex variable
properties hold:
\[
  s_1+s_2 = (\sigma_1+j\omega_1)+(\sigma_2+j\omega_2)
           = (\sigma_1+\sigma_2)+j(\omega_1+\omega_2),
\]
and
\[
  s_1 s_2 = (\sigma_1+j\omega_1)(\sigma_2+j\omega_2)
           = (\sigma_1\sigma_2-\omega_1\omega_2)+j(\sigma_1\omega_2+\sigma_2\omega_1).
\]

A complex variable~$s$ can be geometrically considered as a vector with a length (also
known as magnitude or modulus) and direction (also known as argument or angle) in
either radians or degrees. The magnitude of a complex variable~$s$ is obtained using
the formula
\[
  |s| = \sqrt{\sigma^2+\omega^2},
\]
while the argument of the complex variable~$s$ is obtained using the formula
\[
  \arg s = \tan^{-1}\!\frac{\omega}{\sigma}
\]
in radians for $\omega,\sigma>0$.

The magnitude of the product of two complex numbers $s_1$ and $s_2$ is the product of
their magnitudes, i.e.\ $|s_1 s_2|=|s_1|\,|s_2|$. The argument of the product of two
complex numbers is the sum of their arguments, i.e.\
$\arg(s_1 s_2)=\arg s_1+\arg s_2$.

%--------------------------------------------------------------------------
\section{Roots of Complex Variables}
%--------------------------------------------------------------------------

One of the most important formulas by de~Moivre is
\begin{equation}\label{eq:demoivre}
  \sqrt[n]{s}
  = \sqrt[n]{r}\left[
      \cos\!\left(\frac{\varphi+2k\pi}{n}\right)
      + j\sin\!\left(\frac{\varphi+2k\pi}{n}\right)
    \right],
\end{equation}
for $0\le k\le n-1$, and where $r=|s|=\sqrt{\sigma^2+\omega^2}$ and
$\varphi=\arg s=\tan^{-1}(\omega/\sigma)$ in radians are the magnitude and argument
of~$s$, respectively. This property is very important especially when $n>1$ because
it allows us to evaluate the complex variable with a fractional exponent, i.e.\
$s^{1/n}$, which is predominant in fractional-order systems.

An example of a fractional-order transfer function is
\[
  Z(s) = \frac{V(s)}{I(s)} = \frac{1}{s^{\alpha}C} = \frac{10{,}000}{s^{1/2}},
\]
which is an example of a fractional-order capacitor with pseudo-capacitance
$C=1/10{,}000=100\;\mu\text{F}/\!\sqrt{\text{sec}}$. The reader may refer to~\cite{westerlund1994}
on discussions about capacitor theory.

%--------------------------------------------------------------------------
\section{Results}
%--------------------------------------------------------------------------

%----------------------------------
\subsection{Preliminary Case: $s^{\alpha}=(\sigma+j\omega)^{\alpha}$ where $0<\alpha<1$
  and $\sigma,\omega>0$}
%----------------------------------

Since $0<\alpha<1$, it is assumed that the exponent of the complex variable~$s$ is a
fraction. For simplicity, let us assign $n=1/\alpha$ so we can easily use de~Moivre's
formula in~\eqref{eq:demoivre}.

The magnitude of $s=\sigma+j\omega$ is $r=\sqrt{\sigma^2+\omega^2}$ while its argument
is $\varphi=\arg(s)=\arg(\sigma+j\omega)=\tan^{-1}(\omega/\sigma)$.

Through de~Moivre's formula, the following equation is obtained:
\[
  s^{\alpha} = s^{1/n} = (\sigma+j\omega)^{1/n}
  = \sqrt[n]{r}\left[
      \cos\!\left(\frac{\varphi+2k\pi}{n}\right)
      + j\sin\!\left(\frac{\varphi+2k\pi}{n}\right)
    \right],
\]
for $0\le k\le n-1$. Using the values of $r$ and $\varphi$ in the equation results in
\[
  \sqrt[n]{\sigma+j\omega}
  = \bigl(\sigma^2+\omega^2\bigr)^{1/2n}
    \left[
      \cos\frac{\tan^{-1}(\omega/\sigma)+2k\pi}{n}
      + j\sin\frac{\tan^{-1}(\omega/\sigma)+2k\pi}{n}
    \right],
\]
and since $n=1/\alpha$, the equation can be rearranged as
\begin{equation}\label{eq:general}
  (\sigma+j\omega)^{\alpha}
  = \bigl(\sigma^2+\omega^2\bigr)^{\alpha/2}
    \left[
      \cos\alpha\!\left(\tan^{-1}\frac{\omega}{\sigma}+2k\pi\right)
      + j\sin\alpha\!\left(\tan^{-1}\frac{\omega}{\sigma}+2k\pi\right)
    \right],
\end{equation}
for $0\le k\le n-1$.

By assuming $k=0$, Equation~\eqref{eq:general} can be rewritten as
\begin{equation}\label{eq:general_k0}
  (\sigma+j\omega)^{\alpha}
  = \bigl(\sigma^2+\omega^2\bigr)^{\alpha/2}
    \left[
      \cos\alpha\!\left(\tan^{-1}\frac{\omega}{\sigma}\right)
      + j\sin\alpha\!\left(\tan^{-1}\frac{\omega}{\sigma}\right)
    \right].
\end{equation}

Although this result may not usually be used for systems analysis because we only
normally use the imaginary part of the complex variable in both the Laplace and Fourier
transforms, this result is still beneficial to easily derive the results in the next cases.

%----------------------------------
\subsection{Case~1: $s^{\alpha}=(j\omega)^{\alpha}$ where $0<\alpha<1$ and $\omega>0$}
%----------------------------------

This is a case where the real part is zero, $\sigma=0$. By using the previous result in
Equation~\eqref{eq:general_k0}, and letting $\sigma=0$, we can obtain another
relationship:
\[
  (0+j\omega)^{\alpha}
  = \bigl(0^2+\omega^2\bigr)^{\alpha/2}
    \left[\cos\alpha\!\left(\tan^{-1}(\omega/0)\right)
          +j\sin\alpha\!\left(\tan^{-1}(\omega/0)\right)\right]
  = \omega^{\alpha}\left[\cos\alpha(\pi/2)+j\sin\alpha(\pi/2)\right].
\]
The expression can then be rewritten as
\begin{equation}\label{eq:case1}
  (j\omega)^{\alpha}
  = \omega^{\alpha}\left[\cos\frac{\alpha\pi}{2}+j\sin\frac{\alpha\pi}{2}\right].
\end{equation}

It is straightforward to get the magnitude of this expression. Since the magnitude can
be expressed as a product of individual magnitudes, we can write
\[
  |(j\omega)^{\alpha}|
  = |\omega^{\alpha}|\;\left|\cos\frac{\alpha\pi}{2}+j\sin\frac{\alpha\pi}{2}\right|.
\]
The left multiplicand is a pure real variable, therefore $|\omega^{\alpha}|=\omega^{\alpha}$.
The right multiplicand is a complex variable, so we use the Pythagorean theorem to get
its magnitude:
$\sqrt{\cos^2(\alpha\pi/2)+\sin^2(\alpha\pi/2)}$.
However, the well-known trigonometric identity $\cos^2\theta+\sin^2\theta=1$ renders the
entire square root term equal to unity. The magnitude of Equation~\eqref{eq:case1} then
simply becomes
\begin{equation}\label{eq:case1_mag}
  |(j\omega)^{\alpha}| = \omega^{\alpha}.
\end{equation}

To get the argument of $(j\omega)^{\alpha}$, we note that the argument of the product
of two multiplicands is the sum of their arguments:
\[
  \arg(j\omega)^{\alpha}
  = \arg\omega^{\alpha}
    + \arg\!\left(\cos\frac{\alpha\pi}{2}+j\sin\frac{\alpha\pi}{2}\right).
\]
The argument of $\omega^{\alpha}$, which we can rewrite as $\omega^{\alpha}+j\cdot 0$, is
$\tan^{-1}(0/\omega^{\alpha})=0$. The argument of
$\cos(\alpha\pi/2)+j\sin(\alpha\pi/2)$ is
\[
  \tan^{-1}\frac{\sin(\alpha\pi/2)}{\cos(\alpha\pi/2)}
  = \tan^{-1}\tan\frac{\alpha\pi}{2}
  = \frac{\alpha\pi}{2}.
\]
Therefore,
\begin{equation}\label{eq:case1_arg}
  \arg(j\omega)^{\alpha} = \frac{\alpha\pi}{2}
\end{equation}
in radians.

%----------------------------------
\subsection{Case~2: $a s^{\alpha}+b = a(j\omega)^{\alpha}+b$ where $0<\alpha<1$,
  $\omega>0$, and $a,b>0$}
%----------------------------------

This case is usually found in fractional-order transfer functions. The expression can be
obtained by adding $b>0$ and multiplying $a>0$ in $(j\omega)^{\alpha}$ from
Equation~\eqref{eq:case1}. The resulting expression becomes
\[
  a(j\omega)^{\alpha}+b
  = a\omega^{\alpha}\left[\cos\frac{\alpha\pi}{2}+j\sin\frac{\alpha\pi}{2}\right]+b.
\]
It is necessary to rearrange the terms to cluster the real and imaginary components
separately:
\begin{equation}\label{eq:case2}
  a(j\omega)^{\alpha}+b
  = \left(b+a\omega^{\alpha}\cos\frac{\alpha\pi}{2}\right)
    + j\,a\omega^{\alpha}\sin\frac{\alpha\pi}{2},
\end{equation}
because such arrangement is beneficial when getting the magnitude and argument of
Equation~\eqref{eq:case2}.

The magnitude response of Equation~\eqref{eq:case2} is obtained as
\begin{equation}\label{eq:case2_mag}
  |a(j\omega)^{\alpha}+b|
  = \sqrt{b^2 + a^2\omega^{2\alpha} + 2ab\omega^{\alpha}\cos\frac{\alpha\pi}{2}},
\end{equation}
while the argument is obtained as
\begin{equation}\label{eq:case2_arg}
  \arg\bigl(a(j\omega)^{\alpha}+b\bigr)
  = \tan^{-1}\frac{a\omega^{\alpha}\sin(\alpha\pi/2)}{b+a\omega^{\alpha}\cos(\alpha\pi/2)}.
\end{equation}

%--------------------------------------------------------------------------
\section{Summary}
%--------------------------------------------------------------------------

We now summarize the results in Table~\ref{tab:summary} for the reader's convenience.

\begin{table}[ht]
\centering
\caption{Summary of widely-used results to aid fractional-order systems analysis.}
\label{tab:summary}
\renewcommand{\arraystretch}{2.0}
\begin{tabular}{@{} p{3.8cm} p{9.5cm} @{}}
\toprule
\multicolumn{2}{l}{\textbf{Case~I:} $(j\omega)^{\alpha}$} \\
\midrule
Expression &
  $(j\omega)^{\alpha} = \omega^{\alpha}\bigl[\cos(\alpha\pi/2)+j\sin(\alpha\pi/2)\bigr]$ \\
Magnitude &
  $|(j\omega)^{\alpha}| = \omega^{\alpha}$ \\
Argument [radians] &
  $\arg(j\omega)^{\alpha} = \alpha\pi/2$ \\
\midrule
\multicolumn{2}{l}{\textbf{Case~II:} $a(j\omega)^{\alpha}+b$} \\
\midrule
Expression &
  $a(j\omega)^{\alpha}+b = \bigl(b+a\omega^{\alpha}\cos(\alpha\pi/2)\bigr)
    + j\,a\omega^{\alpha}\sin(\alpha\pi/2)$ \\
Magnitude &
  $|a(j\omega)^{\alpha}+b|
    = \sqrt{b^2+a^2\omega^{2\alpha}+2ab\omega^{\alpha}\cos(\alpha\pi/2)}$ \\
Argument [radians] &
  $\arg\bigl(a(j\omega)^{\alpha}+b\bigr)
    = \tan^{-1}\!\dfrac{a\omega^{\alpha}\sin(\alpha\pi/2)}{b+a\omega^{\alpha}\cos(\alpha\pi/2)}$ \\
\bottomrule
\end{tabular}
\end{table}

Case~I is a simple case where the magnitude and argument of the term $(j\omega)^{\alpha}$
are easily determined through straightforward trigonometry. Case~II, on the other hand, is a
case mostly found in many fractional calculus applications, especially in engineering.
Having $a=1$ and $b=0$ for Case~II makes it equivalent to Case~I.

%--------------------------------------------------------------------------

\end{document}